\documentclass[%
 aip,
 amsmath,amssymb,
 reprint,%
]{revtex4-1}

\usepackage{graphicx}
\usepackage{dcolumn}
\usepackage{bm}
\usepackage{braket}
\usepackage{xcolor}

\usepackage[utf8]{inputenc}
\usepackage[T1]{fontenc}
\usepackage{mathptmx}
\usepackage{etoolbox}
\usepackage{subfiles}
\usepackage{comment}

\let\vec\mathbf

\makeatletter
\def\@email#1#2{%
 \endgroup
 \patchcmd{\titleblock@produce}
  {\frontmatter@RRAPformat}
  {\frontmatter@RRAPformat{\produce@RRAP{*#1\href{mailto:#2}{#2}}}\frontmatter@RRAPformat}
  {}{}
}%
\makeatother

\begin{document}

\preprint{AIP/123-QED}

\title{Integrating Magnons for Quantum Information}
\author{Zhihao Jiang}
\thanks{Zhihao Jiang and Jinho Lim made equal contributions to preparing the first draft. The whole manuscript was prepared and reviewed by all authors.}
\affiliation{Department of Materials Science and Engineering, University of Illinois Urbana-Champaign, Urbana, IL 61801, USA}

\author{Jinho Lim}
\thanks{Zhihao Jiang and Jinho Lim made equal contributions to preparing the first draft. The whole manuscript was prepared and reviewed by all authors.}
\affiliation{Department of Materials Science and Engineering, University of Illinois Urbana-Champaign, Urbana, IL 61801, USA}
\affiliation{Materials Research Laboratory, University of Illinois Urbana-Champaign, Urbana, IL 61801, USA}

\author{Yi Li}
\affiliation{Materials Science Division, Argonne National Laboratory, Lemont, IL 60439, USA}

\author{Wolfgang Pfaff}
\affiliation{Materials Research Laboratory, University of Illinois Urbana-Champaign, Urbana, IL 61801, USA}
\affiliation{Department of Physics, University of Illinois Urbana-Champaign, Urbana, IL 61801, USA}

\author{Tzu-Hsiang Lo}
\affiliation{Department of Materials Science and Engineering, University of Illinois Urbana-Champaign, Urbana, IL 61801, USA}
\affiliation{Materials Research Laboratory, University of Illinois Urbana-Champaign, Urbana, IL 61801, USA}

\author{Jiangchao Qian}
\affiliation{Department of Materials Science and Engineering, University of Illinois Urbana-Champaign, Urbana, IL 61801, USA}
\affiliation{Materials Research Laboratory, University of Illinois Urbana-Champaign, Urbana, IL 61801, USA}

\author{André Schleife}
\affiliation{Department of Materials Science and Engineering, University of Illinois Urbana-Champaign, Urbana, IL 61801, USA}
\affiliation{Materials Research Laboratory, University of Illinois Urbana-Champaign, Urbana, IL 61801, USA}
\affiliation{National Center for Supercomputing Applications, University of Illinois Urbana-Champaign, Urbana, IL 61801, USA}

\author{Jian-Min Zuo}
\affiliation{Department of Materials Science and Engineering, University of Illinois Urbana-Champaign, Urbana, IL 61801, USA}
\affiliation{Materials Research Laboratory, University of Illinois Urbana-Champaign, Urbana, IL 61801, USA}

\author{Valentine Novosad}
\affiliation{Materials Science Division, Argonne National Laboratory, Lemont, IL 60439, USA}

\author{Axel Hoffmann}
\thanks{For any correspondence, please send to axelh@illinois.edu.}
\affiliation{Department of Materials Science and Engineering, University of Illinois Urbana-Champaign, Urbana, IL 61801, USA}
\affiliation{Materials Research Laboratory, University of Illinois Urbana-Champaign, Urbana, IL 61801, USA}
\affiliation{Department of Physics, University of Illinois Urbana-Champaign, Urbana, IL 61801, USA}
\email{axelh@illinois.edu}


\begin{abstract}
Magnons, the quanta of collective spin excitations in magnetically ordered materials, have distinct properties that make them uniquely appealing for quantum information applications. They can have ultra-small wavelengths down to the nanometer scale even at microwave frequencies. They can provide coupling to a diverse set of other quantum excitations, and their inherently gyrotropic dynamics forms the basis for pronounced non-reciprocities. In this article we discuss what the current research challenges are for integrating magnetic materials into quantum information systems and provide a perspective on how to address them. 
\end{abstract}

\maketitle

\section{\label{sec:introduction}Introduction}

The coherent manipulation and control of individual quantum states is at the heart of many new approaches to ultimate-precision sensors, securely encrypted communication, and beyond classical computation~\cite{Kurizki2015}. In order to pursue this goal a wide variety of fundamental excitations are being considered, each with their own distinct advantages and drawbacks. For example, microwave photons are often used in superconducting quantum systems for their ease of being incorporated into well-defined circuitry for complex interactions, but require very low temperatures to operate in the single quantum limit~\cite{Blais2021}. On the other hand, optical photons are robust against decoherence even at room temperature and therefore well suited for long-distance transportation of quantum information~\cite{Wang2020}. These two examples highlight the fundamental dilemma that is inherent to many quantum systems, namely the desire to maintain long spatial and temporal coherence, while at the same time having the ability to coherently manipulate on demand quantum states with high fidelity. These two requirements for functional quantum systems are often antagonistic due to their underlying physics. Thus, it is expected that for the realization of large-scale quantum information systems many different quantum excitations will be integrated to incorporate their specific advantages.

For practical implications of quantum measurements or computation approaches, one often desires to have a well-defined system with only two distinct and isolated states. This is naturally achieved by spin-$\frac{1}{2}$ systems, where the individual spin up and down states can be controlled by intrinsic and extrinsic magnetic fields. Given the relatively long coherence times of both nuclear and electronic spins it is not surprising that some of the early implementations of quantum computation were performed by using nuclear spins~\cite{Gershenfeld1997,Vandersypen2001} and that electron spins at atomically sized defects provide some of the most promising new sensing technologies~\cite{Schirhagl2014}. However, the long coherence is due to the relatively weak coupling of individual spins, which also makes it challenging to integrate them with more complex hybrid quantum systems. This challenge can be overcome by using larger spin ensembles, or ultimately strongly interacting spin systems, such as magnetically ordered materials, which provide the highest spin-density in nature. 

In these magnetically ordered systems the fundamental excitations are given by magnons. The high spin-density associated with magnetically ordered materials enables strong coupling to magnons. This of course gives rise to challenges with respect to their coherence time, but nevertheless coherent interactions with magnons at the single quantum level have already been demonstrated~\cite{Lachance2020,Tabuchi2015,Xu2022}. As discussed in detail in this perspective, magnons provide several unique properties that can enhance and broaden the current technology for quantum information processing and engineering~\cite{Li2020,Awschalom2021,Yuan2022}.

\begin{figure*}
    \centering
    \includegraphics[width=0.8\textwidth]{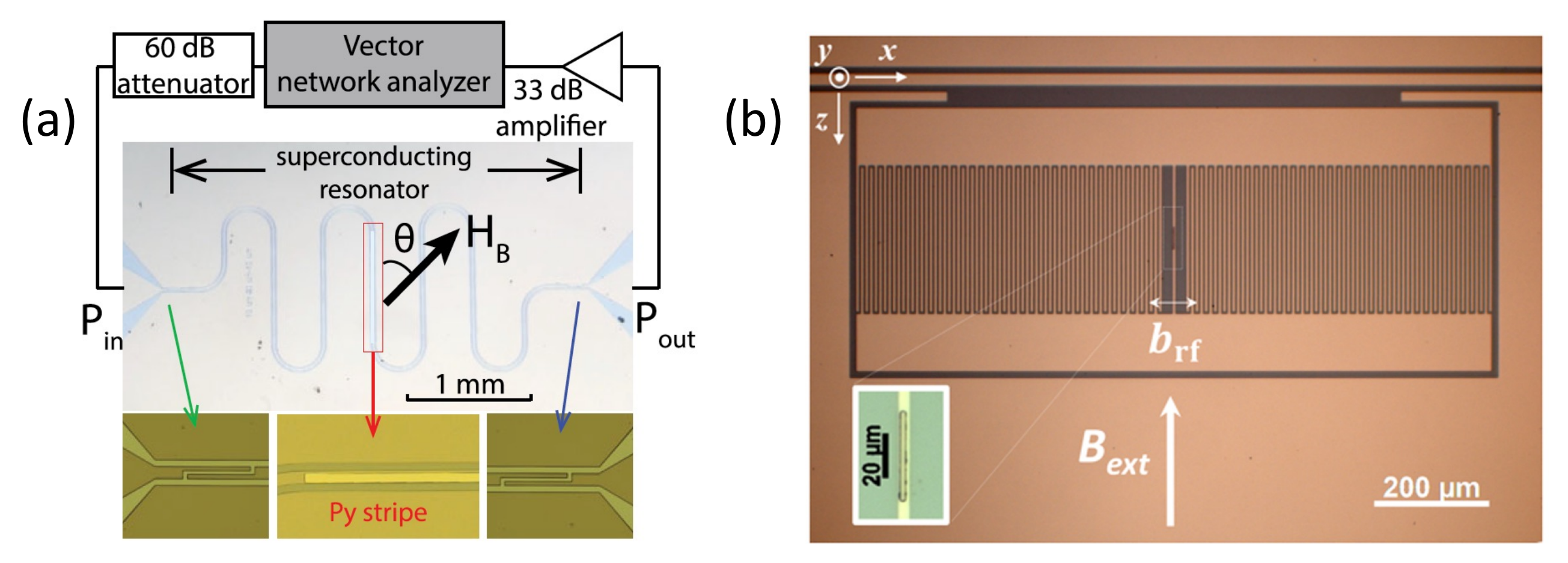}
    \caption{Examples of magnetic elements integrated with superconducting resonators based on (a) striplines or (b) lumped element structures. Both structures demonstrated strong coupling between magnons and microwave photons. Adapted from Refs.~\cite{Li2019} and~\cite{Hou2019}.}
    \label{fig_intro:on_chip_coupling}
\end{figure*}

Among the distinct advantages of magnons is that they can couple to a wide variety of other excitations, including microwave and optical photons, phonons, other magnons, and individual spin systems. Thus, magnons are well-suited to act as a transducer between different quantum excitations, e.g., between microwave and optical photons~\cite{Zhang2016}. As shown in Fig.~\ref{fig_intro:on_chip_coupling}, magnetic elements can be readily integrated into planar, high-quality, superconducting microwave circuitry~\cite{Li2019, Hou2019}. Compared to other microwave excitations (photons or phonons), the wavelengths of magnons can be significantly shorter, which makes magnetic elements more suitable for miniaturization and their performance less restricted by their dimensions.  At the same time the use of magnetic materials in spintronic applications~\cite{Hoffmann2015,Dieny2020} has established advanced approaches for integrating them even with complex nanoscale devices. But one of the more distinct advantages of magnetic systems is that the time-reversal breaking of the magnetic order and the corresponding gyrotropic magnetization dynamics has an intrinsic chirality, which results in pronounced non-reciprocities. Therefore, magnetic materials are key to many directional microwave devices, such as isolators or circulators. Such unidirectional microwave propagation is highly desirable for quantum devices, since it provides noise isolation, but also can be utilized for novel chiral quantum systems~\cite{Lodahl2017}.

The goal of this perspective is to provide an overview of the current challenges and opportunities for integrating magnetic materials into more sophisticated quantum information systems. We first will discuss in Section~\ref{sec:damping} magnetic damping, which is the key materials parameter determining the coherence of magnons. Thus, reducing magnetic damping is paramount for utilizing magnetic materials and we will discuss several strategies for doing so in hybrid quantum devices. We note that reducing damping also has become a key challenge for current spintronics devices~\cite{Dieny2020}. Next, we will discuss in Section~\ref{sec:nonreciprocal} pathways for engineering non-reciprocities into miniaturized microwave devices that can potentially be integrated directly onto a chip. This ability has implications well beyond just quantum devices, since the non-reciprocity is also required for many classical microwave applications. Lastly, in Section~\ref{sec:magnon-qis} we will discuss two aspects of hybrid magnon devices directly relevant for quantum state transfers. First, we will discuss how different magnon systems can be remotely coupled via microwave photons and subsequently we discuss the new opportunities that arise from coupling different quantum systems via non-reciprocal propagating magnons.

\section{\label{sec:damping}Materials optimization for low magnetic damping}

Maintaining the coherence of quantum states is essential for quantum information processing. To achieve this in a hybrid quantum system, strong coupling between constitute subsystems as well as low decoherence in each subsystem is desirable such that a high cooperativity can be obtained~\cite{Li2020}. In the scope of this paper we focus on damping of magnons in magnetic materials which are essential components of hybrid magnonics. Magnons are quanta of spin waves that are collective excitations of spin precession in a magnetically ordered material. Damping is the relaxation of this precession motion towards the equilibrium spin direction due to energy loss. This leads to the observed field (or frequency) linewidth $\Delta H$ of the magnon mode in, for example, ferromagnetic resonance (FMR) measurements. A widely used empirical formula for $\Delta H$ is~\cite{Rezende2020} 

 \begin{align}
    \Delta H = \Delta H_0 + \frac{2\pi f \alpha}{\gamma} \label{deltaH}
\end{align}
where the first term $\Delta H_0$ is called the inhomogeneous broadening and the second term includes the Gilbert damping derived from the Landau-Lifshitz-Gilbert equation~\cite{Gilbert2004}
\begin{align}
    \frac{d\vec M}{dt} = -\gamma \mu_0 \vec M \times \vec H + \frac{\alpha \vec M}{M} \times \frac{d\vec M}{dt}. \label{LLG}
\end{align}
Here $f$ is the resonant frequency, $\gamma$ is the gyromagnetic ratio, $\alpha$ is called the Gilbert damping parameter, $\mu_0$ is the Bohr magneton, $\vec M$ is the magnetization, and $\vec H$ is an effective magnetic field comprised of both externally applied fields, as well as internal magnetic fields which can originate from exchange coupling, crystalline anisotropies, and other possible contributions.   

This empirical expression \eqref{deltaH} can fit many experimental measurements well, but not all especially in situations when non-Gilbert damping becomes important~\cite{Jermain2017,Maier-Flaig2017}. A microscopic description of magnon damping is generally complicated due to a wide variety of possible contributions. Moreover, damping also depends on the shape and quality of the sample, and the temperature in a non-trivial way~\cite{Jermain2017,Maier-Flaig2017,Khodadadi2020}. There is no universal rule for optimizing magnon damping in all magnetic materials. In this perspective we will discuss this for magnetic insulators, specifically yttrium iron garnet ($\mathrm{Y_3Fe_5O_{12}}$, YIG), and magnetic conductors separately, as they are dominated by different damping mechanisms. Consequently, strategies for optimization are also suggested to be different for these two categories of materials for particularly thin-film samples at cryogenic temperatures due to their potential applications in quantum information science.

\subsection{Yttrium iron garnet, a ferrimagnetic insulator}

The ferrimagnetic insulator yttrium iron garnet ($\mathrm{Y_3Fe_5O_{12}}$, YIG) is so far the best low-damping magnetic material, which has $\alpha \sim 10^{-5}$ for bulk samples~\cite{Glass1976}. The low damping is due to two important factors. First, magnon scattering to conduction electrons is absent in the insulator. Second, the magnetic ion $\mathrm{Fe^{3+}}$ has nearly zero orbital angular momentum. Therefore, the spin-orbit coupling (SOC) is weak, and thus the spin-lattice relaxation time is long. The main contribution to damping in YIG is magnon-magnon scattering. Three-magnon processes induced by dipolar interactions dominate the damping of magnons with smaller wave-number $k$, while four-magnon processes induced by both dipolar interactions and exchange interactions become more significant for large-$k$ modes~\cite{Rezende2020}. For the uniform precession mode (the Kittel mode), i.e., $k=0$, these two contributions are also small, leading to extremely low magnetic damping observed in FMR measurements~\cite{Glass1976}.

The intrinsic damping in ideal YIG is very small. However, in some situations, experimentally observed damping can be larger. This is particularly significant in YIG thin films whose damping parameter is typically an order of magnitude larger compared to bulk samples~\cite{Schmidt2020}. One important damping source in YIG thin films comes from surface imperfections~\cite{Sun2012,Onbasli2014,Howe2015,Tang2016}. Defects on sample surfaces or its interfaces with other materials cause two-magnon scattering, which can increase the FMR linewidth. A theoretical description of this mechanism was proposed by Arias and Mills in 1999~\cite{Arias1999}, where the Kittel mode can be scattered to several degenerate modes of similar energy and small wave-numbers by surface defect potentials. This damping effect is enhanced for thinner YIG films as the surface to volume ratio is increased. Many experimental works have confirmed this by an observed trend of increasing damping with decreasing film thickness~\cite{Sun2012, Kelly2013, Onbasli2014}. Meanwhile, smaller damping has been observed in films with smoother surfaces~\cite{Sun2012,Onbasli2014,Howe2015,Tang2016}.  

\begin{figure}
    \centering
    \includegraphics[width=8.6cm]{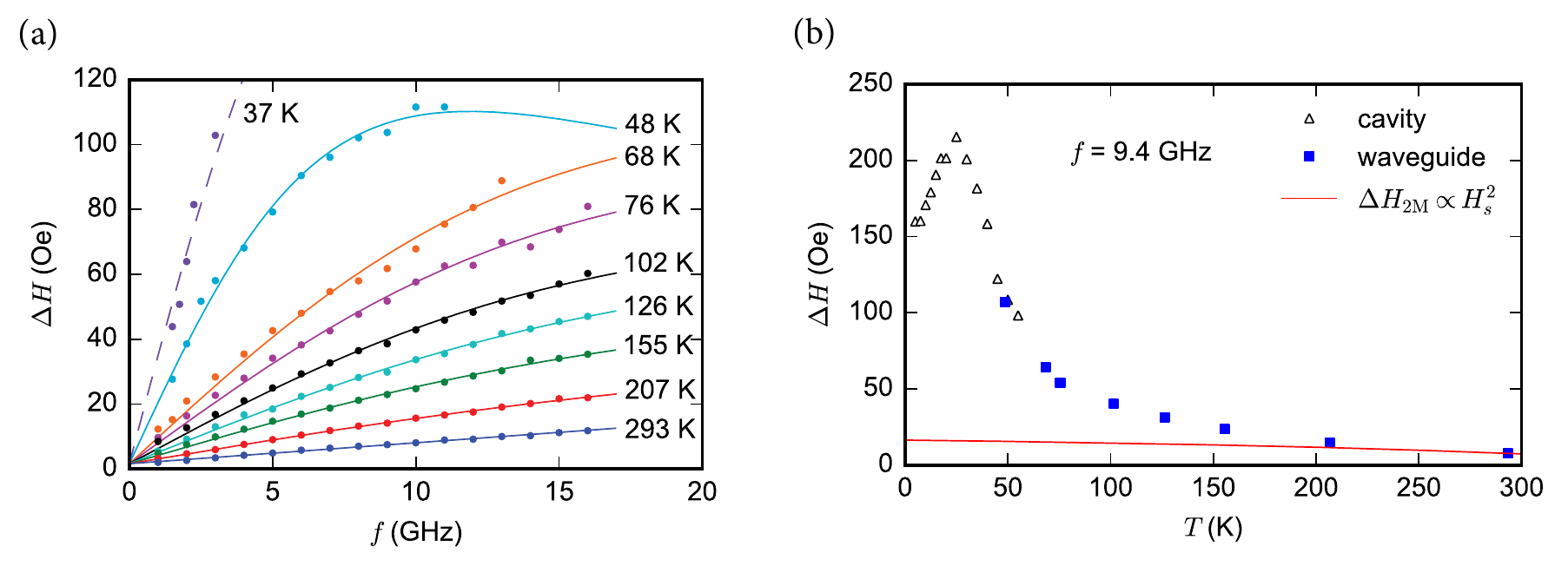}
    \caption{(a) Field linewidth $\Delta H$ versus frequency $f$ for a YIG thin film on a gadolinium gallium garnet substrate at different temperatures. (b) Temperature dependence of the field linewidth $\Delta H$ at fixed frequency $f=9.4\text{ GHz}$. (a) and (b) are adapted from Ref.~\cite{Jermain2017}.}
    \label{fig_damping_1}
\end{figure}

Magnon damping in YIG films at low temperatures has very intriguing behaviors. Jermain et al. studied magnon damping in 15-nm YIG film from room temperature ($297\text{ K}$) to $\sim 10 \text{ K}$ and had several interesting observations~\cite{Jermain2017}. As the temperature decreases, the slope of the FMR linewidth $\Delta H$ versus frequency $f$ is increasing, indicating a higher damping parameter $\alpha$ for lower temperatures [see Fig.~\ref{fig_damping_1} (a)]. Meanwhile, as the temperature goes below $\sim 50\text{ K}$, the linewidth $\Delta H$ strongly deviates from the linear dependence on the frequency $f$. This implies that non-Gilbert damping starts to become important. Moreover, for a fixed frequency of $9.4\text{ GHz}$, the linewidth $\Delta H$ shows a non-monotonic variation with decreasing temperature $T$. A peak was observed near 25 K, where the value of $\Delta H$ was 28 times larger than at room temperature [see Fig.~\ref{fig_damping_1} (b)]. Such a ``temperature-peak" character was also reported in Ref.~\cite{Maier-Flaig2017} for a YIG sphere. In both studies, the authors explained the increased low-temperature damping by slow-relaxing impurities~\cite{Spencer1959, Dillon1959, Sparks1961, Spencer1961} that are perhaps rare earth or $\mathrm{Fe^{2+}}$ impurities introduced during film growth.

Another important damping source for YIG thin films at low temperatures comes from the substrate. High quality YIG films are usually grown via epitaxy on gadolinium gallium garnet ($\mathrm{Gd_3Ga_5O_{12}}$, GGG) for its close lattice matching with YIG. At low temperatures $\lessapprox 70$ K, GGG exhibits paramagnetic behavior and couples magnetically to YIG, acting as a damping source for magnons in YIG~\cite{Danilov1989, Danilov2002, Mihalceanu2018}. In the work of S. Kosen et al.~\cite{Kosen2019}, magnon damping in YIG films at millikelvin (mK) temperatures with and without GGG substrate has been investigated. Significantly increased magnon linewidth was observed for YIG/GGG, while substrate-free YIG can possibly achieve lower damping than the room-temperature value. Moreover, they observed that damping over $1\text{ K}$ shows a similar "temperature-peak" behavior due to slow-relaxing impurities, while below $1\text{ K}$ the damping saturates, which is attributed to two-level fluctuators. Similar effects of impurities damping and substrate damping were also reported in Ref.~\cite{Mihalceanu2018}. In both works, the thickness of films is on the micrometer level so the two-magnon scattering effect is likely to be smaller than for films with sub-micrometer thickness.      

Improving sample perfection and minimizing substrate effects with more advanced experimental techniques is therefore a main direction for future efforts to reduce damping further. Over the last decade, many efforts have been made to fabricate YIG thin films with better crystal quality and, especially, smoother surfaces, via different growing methods and pre- or post-processing strategies. A detailed review on these can be found in Ref.~\cite{Schmidt2020}. Particular attentions are also paid on the atomic structure of the sample surface or its interface with the substrate. Several studies have observed the presence of a Fe-loss and Ga-rich transition layer between YIG and GGG~\cite{Mitra2017,Cooper2017,Quarterman2022}. This transition layer, also known as the magnetically dead layer has a thickness of several nanometers, depending on the growth process. Near the top of YIG films, Song et al. have reported oxygen deficiency within a thin surface layer of a few nanometers and the presence of a disordered layer~\cite{Song2015}. How these surface structural effects influence the damping and the magnetization of YIG films remains an interesting research topic. One the other hand, when going to cryogenic temperatures, relaxation due to impurities such as rare-earth elements and $\mathrm{Fe^{2+}}$ ions, and relaxation due to substrate effects from GGG, start to dominate~\cite{Jermain2017, Maier-Flaig2017, Kosen2019, Mihalceanu2018}. Growing YIG films with high chemical purity should be pursued. At a few Kelvin or even lower temperatures, coupling to the substrate GGG is a main source of damping. Alternative substrates or free-standing YIG are potential solutions to be considered. Silicon (SI) has been tried as another substrate for YIG, which, at room temperature, gives apparently larger damping than YIG/GGG~\cite{Balinskiy2017, Delgado2018}. At cryogenic temperatures, it is not clear whether YIG/Si could possibly outperform YIG/GGG since Si does not magnetically couple to YIG. Trempler et al.~\cite{Trempler2020} presented a process which can transfer single-crystal yttrium-iron-garnet microstructures from GGG to other kinds of substrates and reported low damping on the order of $10^{-4}$ at 5 K. Recently, Guo et al.~\cite{Guo2022} have tried embedding a thin diamagnetic $\mathrm{Y_3Sc_{2.5}Al_{2.5}O_{12}}$ (YSAG) bilayer between YIG and GGG to eliminate the exchange coupling between both. As a result, they obtained much reduced damping at 2–5 K in this YIG/YSAG/GGG structure compared to YIG/GGG. In a subsequent work from the same group, they grew YIG films on a new diamagnetic $\mathrm{Y_3Sc_2Ga_3O_{12}}$ (YSGG) single-crystal substrate and observed smaller damping than for YIG/GGG~\cite{Guo2022arXiv}.

\subsection{Metallic ferromagnets}
\begin{figure}
    \centering
    \includegraphics[width=8.6cm]{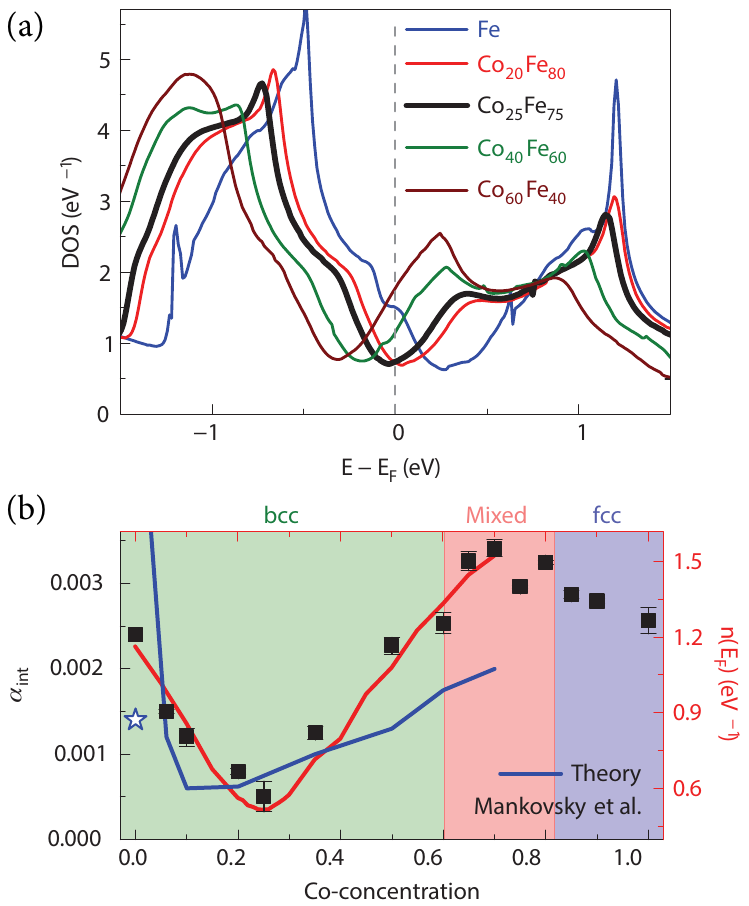}
    \caption{(a) Electronic density of states (EDOS) of cobalt-iron (Co-Fe) alloy with different Co concentrations. (b) Experimentally measured intrinsic damping of Co-Fe alloys with different Co concentrations, compared to theoretical calculations from Mankovsky et al~\cite{Mankovsky2013}. The change of the electronic density of states (EDOS) is also shown in the plot. (a) and (b) are adapted from Ref.~\cite{Schoen2016}.}
    \label{fig_damping_2}
\end{figure}

Ferromagnetic metals and alloys are another category of promising magnetic materials for hybrid magnonics applications, especially since their integration into heterostructures and devices is significantly easier than for insulating materials such as YIG. However, these metallic materials, such as permalloy, have generally larger magnetic damping than insulators due to the presence of conduction electrons which contribute as a dominant damping pathway. Nevertheless, an ultra low damping of the order of $10^{-4}$ was observed in $10$-nm thick polycrystalline cobalt-iron (CoFe) alloy films with 25\% Co concentration~\cite{Schoen2016}. Such a low damping is attributed to reduced magnon-electron scattering indicated by a sharp minimum of the electronic density of states (EDOS) at the Fermi level, as illustrated in Figs.~\ref{fig_damping_2}(a) and (b). Comparably low damping was also reported in a single-crystal $\mathrm{Co_{25}Fe_{75}}$ sample by Lee et al.~\cite{Lee2017}. These observations suggest the possibility for metallic ferromagnets to have damping properties almost comparable to YIG, while also having the advantage that the magnetization is about one order of magnitude larger, enabling stronger coupling. In order to optimize these materials for low damping the key factor is the electronic structure of the material.  

Theories of magnon-electron damping in metallic ferromagnetic materials have been investigated for quite some time. Two early models developed by Kamberský are the breathing Fermi surface (BFS) model~\cite{Kambersky1970} and the torque correlation (TC) model~\cite{Kambersky1976}. In the BFS model, magnetization precession induces electron-hole (e-h) pairs near the Fermi level $E_f$ by pushing some occupied (unoccupied) states above (below) the Fermi level $E_f$ through spin-orbit interaction. These e-h pairs are damping sources for magnons. In this case, longer lifetime of the e-h pairs is associated with larger magnon damping. This leads to a naively counter-intuitive result that this damping mechanism is enhanced at lower temperatures and in cleaner samples because it is proportional to the electron scattering time $\tau$. Similar results to the BFS model were also derived by Korenman and Prange for explaining the observed increased low-temperature magnetic damping (by that time it was called ``anomalous damping") in nickel and cobalt~\cite{Korenman1972}. The BFS damping mechanism was later recognized as the conductivity-like damping in contrast to the resistivity like damping, which is detailed in the TC model~\cite{Kambersky1976, Kambersky2007, Gilmore2008}. In the TC model, the spin-orbit torque $\Gamma^{-}_{mn}=\braket{m,\vec k|[\sigma^{-}, \hat{H}_\text{SO}]|n, \vec k}$ induces damping. Here $\hat{H}_\text{SO}$ is the Hamiltonian operator of spin-orbit interactions, which should not be confused with the magnetic field. $\ket{n, \vec k}$ is the eigenstate with band index $n$ and wavevector $\vec k$. This spin-orbit torque can be decomposed into the intra-band ($m = n$) part and the inter-band ($m \neq n$) part~\cite{Kambersky2007, Gilmore2008}. While the intra-band part essentially reproduces the same phenomena as in the BFS model, the inter-band part becomes important only when the spectral overlap between different bands becomes large. This can arise from broadening of energy bands by increased temperature or disorder. The inter-band part is therefore considered as resistivity-like damping as it is inverse proportional to the electron scattering time $\tau$ and is enhanced when the temperature is high or when the system is more disordered. These two mechanisms, i.e., intra-band and inter-band scattering, imply an interesting result that magnon damping in metallic materials is \textit{not monotonically} dependent on the temperature or disorder, which both are correlated to the electron scattering time $\tau$. This trend explicitly shows up in first-principles calculations~\cite{Gilmore2007} and is indeed observed in several $3d$ transition metals~\cite{Bhagat1974}. The BFS model and TC model have also been generalized to non-local and anisotropic version by replacing the scalar damping parameter with a damping tensor $\Tilde{\bm{\alpha}}$~\cite{Fahnle2008, Gilmore2010, Thonig2014, Turek2015, Thonig2018, Brataas2008}.              

Magnetic damping at low temperatures is of particular interest here since we are aiming for quantum information science applications. In this low scattering limit, damping is largely dominated by the intra-band scattering mechanism. Theories have demonstrated the explicit dependence of damping on the electronic density of states and spin-orbit interaction~\cite{Gilmore2008}. One simplified but useful formula states $\alpha \sim N(E_F)|\Gamma^-|^2 \tau$, highlighting that the density of states at the Fermi level $N(E_F)$, the spin-orbit coupling (SOC) coupling parameter $\Gamma^- = \braket{[\sigma^{-}, \hat{H}_\text{SO}]}_{E=E_F}$ at the Fermi energy, and the electron scattering time $\tau$ are key factors determining the intra-band damping. Many experiments, combined with first-principles studies, have confirmed that damping can be reduced by minimizing $N(E_F)$~\cite{Schoen2016, Schoen2017, Arora2021, Wei2021}. Recently, Khodadadi et al.~\cite{Khodadadi2020} observed that at 10 K an epitaxial thin film of pure iron with imperfect crystallinity exhibits lower Gilbert damping than the cleaner counterpart. This is again explained by the intra-band conductivity-like damping being proportional to $\tau$ and therefore inversely proportional to lattice disorder. In addition, experimentally measured damping shows giant anisotropy in some materials such as Fe and Co-Fe-(B) thin films on substrates~\cite{LChen2018, Li2019Aniso, Xia2021, LChen2023}. This can be explained by the anisotropy in SOC or the anisotropy in $N(E_F)$ induced by interfacial effects.

Minimizing electron-magnon scattering is the main pathway for reducing magnetic damping in metallic materials. The available phasespace for optimal materials is very broad and includes different magnetic metals and alloys. This situation is very different from the case of YIG being the ``dominant" low-damping material among magnetic insulators. In addition, any factor affecting electronic properties in a material can possibly be considered as a tuning factor for magnon-electron damping. Alloy compositions, temperature, crystal quality, and chemical doping are several factors that can affect magnetic damping in metallic magnets~\cite{Rantschler2007, Mizukami2011, Luo2014, Schoen2016, Khodadadi2020, Arora2021, Wei2021, Wu2022}. Recently, Wang et al.~\cite{JWang2019} found that structural transformations from the crystalline to the amorphous state provide a pathway for reducing magnetic damping in Co-Fe-C alloy films. Therefore, we expect that combined theoretical, computational, and experimental efforts will be pursued for a more efficient exploration of optimal low-damping metallic magnetic materials. Recently emerged studies in two-dimensional (2D) van der Waals magnets and topological magnetic semi-metals may bring new theoretical insight into electronic structure and magnetic dynamics in such novel materials~\cite{Gong2017, Alahmed2021, YZhang2021, Swekis2021, Nikolic2021}. In the computational community, first-principles calculations of intrinsic magnon damping in magnetic metals and alloys have already been done for many years, providing qualitatively good agreement with experiments~\cite{Kunes2002, Gilmore2007, Starikov2010, Liu2011, Ebert2011, Mankovsky2013, Turek2015, Chico2016, Mankovsky2022}. With the developing computational power for more precise materials simulations and new developing techniques such as machine learning, we may expect even more efforts and contributions from this community towards an efficient way of materials discovery and optimization. Nevertheless, experiments are natural required for verifying all theories and computational methods developed, and test the true performance of the optimized materials in real materials systems and devices.

\section{\label{sec:nonreciprocal}Spin-wave nonreciprocity}

Wave nonreciprocity is the phenomenon that waves propagating along direction have a different dispersion relation or amplitude than those propagating along the opposite direction. Some necessary conditions for systems to have wave nonreciprocity are broken time-reversal
symmetry~\cite{Camley1987,Barman2021,Kuss2021,Haldane2008,Verba2019,Di2015SRP,Maznev2013}, or structures having asymmetric features~\cite{Maznev2013,Liang2010}. One example is a system having an array of triangular reflectors where one wave is pointing at a vertex of the triangular reflector and the other wave is pointing at an edge so that two oppositely propagating waves experience different potentials for scattering~\cite{Maznev2013}. Another example is a system having an ordered nonlinear region (left)/filter (right) where a wave passes through the nonlinear region first and is converted to another mode by a nonlinear process and the converted mode passes through the filter, but another wave, which hits the filter first doesn not pass the filter~\cite{Liang2010}. 

Here, we will focus on spin wave nonreciprocity with a linear excitation level in systems with broken time-reversal symmetry corresponding to every ferromagnetic material which inherently have broken time-reversal symmetry due to their net magnetization. Nonreciprocal characters of magnon propagation have been implemented to ferromagnetic systems in several ways as an intrinsic property or an extrinsic property of the system. Implementable intrinsic properties that can support nonreciprocal magnon propagation are interfacial Dzyaloshinskii–Moriya (DM) interaction~\cite{Moon2013,Lan2015,Kuss2020,WangHC2020,Di2015PRL,Liang2022} from layered structures of ferromagnet (FM)/heavy metal (HM), bulk DM interaction with materials whose lattice structures do not have inversion symmetry~\cite{Iguchi2015,Seki2016}, interlayer dipolar interaction between two FM layers~\cite{Kuss2021,Di2015SRP,Odintsov2022,Chen2019,Gallardo2019PRApp,Albisetti2020,Gallardo2019IOP,Gallardo2022}, altered Maxwell boundary condition in the presence of nearby metallic layer~\cite{Lim2018,Trossman2019,Trossman2018,Bongianni1972}, and altered exchange boundary condition (EBC) by using surface anisotropy~\cite{Gladii2016}. These nonreciprocal magnetic systems are expected to provide noise isolation in quantum information systems on the chip level because isolators, used in the conventional quantum information system with microwave photon to prevent backflow of thermal radiation at output line, may be replaced by these nonreciprocal magnetic systems which can be integrated closer to superconducting devices.

Interfacial DM interactions have a term proportional to $\vec S_i \times \vec S_j$ which have different signs at opposite spin wave chirality or equivalently opposite spin wave propagating direction, so that DM interactions generate nondegenerate magnon dispersions between opposite propagation directions. More precisely, there is an additional linear dependence on the wavevector, $k$, in the dispersion, $\omega(k)$~\cite{Moon2013}, so two magnons propagating in opposite directions have different bandwidths and group velocities, thereby they have also different attenuations. Note that since interfacial DM interaction is an interlayer exchange interactions acting on an atomic length-scale, it is effective only for ultra-thin films of typically a few nanometers. Bulk DM interactions can also generate nondegenerate magnon dispersions~\cite{Kataoka2013,Melcher1973} in similar ways with those originating from interfacial DM interactions mentioned above so they can also generate nonreciprocity of magnon propagation. But bulk DM interactions typically require single crystal materials without structural chiral twins. This requirement limits the appeal of such materials for devices based on thin films.

\begin{figure}
    \centering
    \includegraphics[width=8.6cm]{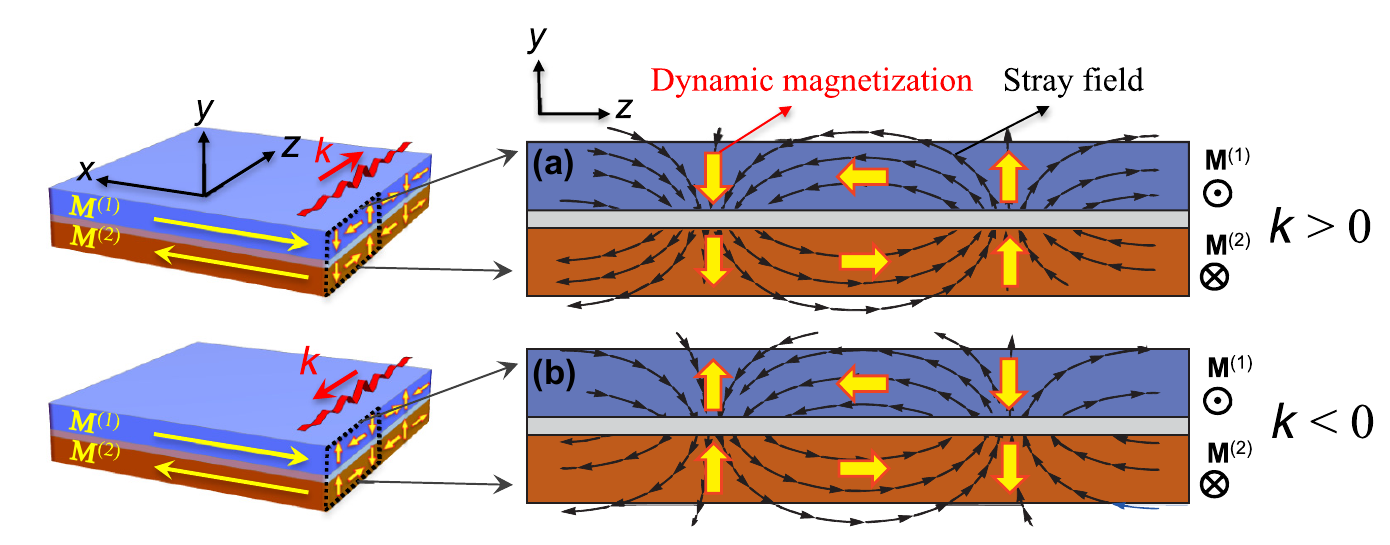}
    \caption{Dynamic dipolar field due to time-varying component of magnetization in the presence of an excited spin wave mode propagating along $z$-axis. The static magnetizations are aligned along $x$-axis and dynamic dipolar field have only $y$- and $z$-component. The dynamic dipolar field with $k_z > 0$ is given at (a) and with $k_z < 0$  is given at (b). Adapted from Ref.~\cite{Gallardo2019PRApp}.}
    \label{fig_nonreciprocal}
\end{figure}

Interlayer dipolar interactions between two FM layers generate terms which are nonsymmetric under wavevector inversion, such that they have a maximum nonreciprocity when the equilibrium magnetization is perpendicular to the magnon wavevector, $\vec M_0 \perp \vec k$~\cite{Gallardo2019PRApp}. There are two possible configurations in the $\vec M_0 \perp \vec k$ geometry which are parallel (ferromagnetic coupling between FM layers; $\vec M_1 \cdot \vec M_2 > 0$) and anti-parallel (antiferromagnetic coupling between FM layers; $\vec M_1 \cdot \vec M_2 < 0$). In the parallel configuration, nonreciprocity due to interlayer dynamic dipolar interaction exists except in a case of $M_1=M_2$ which effectively corresponds to a single layer with double thickness. Nonreciprocity becomes more pronounced with larger $|M_1-M_2|$ values. In the anti-parallel configuration, nonreciprocity always exists and is generally bigger than that in the parallel configuration~\cite{Gallardo2019PRApp}. Fig.~\ref{fig_nonreciprocal} shows two interlayer dipolar fields with two oppositely propagating spin wave modes in the anti-parallel configuration. Since the $y$-component of the dynamic dipolar field and the time-varying component of magnetization are antiparallel in Fig.~\ref{fig_nonreciprocal}(b), the spin wave described in Fig.~\ref{fig_nonreciprocal}(b) has a higher frequency than the spin wave described in Fig.~\ref{fig_nonreciprocal}(a). 

There are a number of ways to engineer either FM or AFM coupling between two FM layers. These are, for example, direct interlayer exchange interaction~\cite{Li2020PRL,Klingler2018}, indirect interlayer exchange interaction with a non-magnetic spacer\footnote{For example, sub-nanometer Iridium or Ruthenium can be used for RKKY interaction.} (such as, RKKY interaction)~\cite{Gallardo2019PRApp,Albisetti2020,Bruno1999,Ishibashi2020}, and shape anisotropy (such as, using a magnonic crystal or grating layer)~\cite{Di2015SRP,Chen2019}. Noticeable nonreciprocity with FM couplings between layers has been observed for multi-layer heterostructures where different saturation magnetizations, $M_s$, are assigned at different layers~\cite{Gallardo2019IOP}. One strong point of using FM coupled systems is that they show nonreciprocity over a wide range of applied magnetic field, i.e., a strong magnetic field does not break FM coupled states when field and magnetization are pointing the same direction, which will simplify applications to magnonic devices. On the other hand, an AFM coupled system can typically maintain its AFM coupled state within a relatively limited range of applied magnetic fields. For instance, AFM coupled systems can have an operating ranges of applied magnetic fields as has been demonstrated for Co nanowires, which maintained their AFM arrangement over 800 Oe~\cite{Di2015SRP,Chen2019}.

An alternative pathway for engineering nonreciprocities is given by controlling the
boundary conditions for chiral surface spin waves, known as Damon-Eshbach modes. In the presence of a metallic layer, the Maxwell boundary condition is altered so that the normal component of the magnetic field, $\vec h_{\perp}(\vec r, t)$, needs to be zero at the FM/Metal interface~\cite{Bongianni1972}. This asymmetric Maxwell boundary condition between two surfaces results in a nondegenerate dispersion of Damon-Eshbach (DE) surface spin waves~\cite{Lim2018}, but in this case the wavevector direction is indirectly related to the altered Maxwell boundary condition. Since DE surface waves propagating in opposite direction excite on opposite surfaces ($\vec k \propto \vec n \times \vec M_0$ where $\vec n$ is a film normal), each DE surface wave is subject to a different Maxwell boundary condition so that they have different dispersion relations.

Similarly, when a magnetic film has an out-of-plane uniaxial surface anisotropy at one surface or different anisotropies at opposite surfaces, each surface has a different EBC or a different spin pinning as one can find in the Rado-Weertman relation~\cite{Rado1959,Labrune1995}. As a result, the dispersion relation of DE surface waves becomes nondegenerate, and in this case also has an implicit relation between wavevector direction and the perturbed EBC. From the same argument used for the Maxwell boundary condition, each of the DE surface waves is subject to different EBCs so that they have different dispersion relation.

In addition to mechanisms for the nonreciprocity described above, it is possible that interlayer direct exchange interactions from exchange biased FM/antiferromagnet (AFM) bilayer system would also be able to create a nonreciprocity. The interlayer exchange interaction results in asymmetric spin pinning at opposite surfaces and therefore surface spin waves propagating on each surface should have different dispersion relation or different amplitude profiles along thickness direction. The interlayer exchange interactions have been broadly studied in the context of various phenomena, for example, exchange bias~\cite{Nogues1999,Nogues2005,Stamps2000}, exchange spring~\cite{Fullerton1999,Hellman2017,Skomski2003}, giant magnetoresistance (GMR)~\cite{Binasch1989}, spin Hall effect~\cite{Saito2021}, etc. But so far AFM/FM interlayer exchange coupling has not been sufficiently studied in terms of magnetization dynamics, especially in the context of spin wave propagation. To study how the interlayer exchange interaction affects spin wave propagation, FM/AFM bilayer structures are adequate systems because they do not have interlayer dipolar interactions that FM1/FM2 bilayer systems have in addition to interlayer exchange interactions.

\begin{figure}
    \centering
    \includegraphics[width=8.6cm]{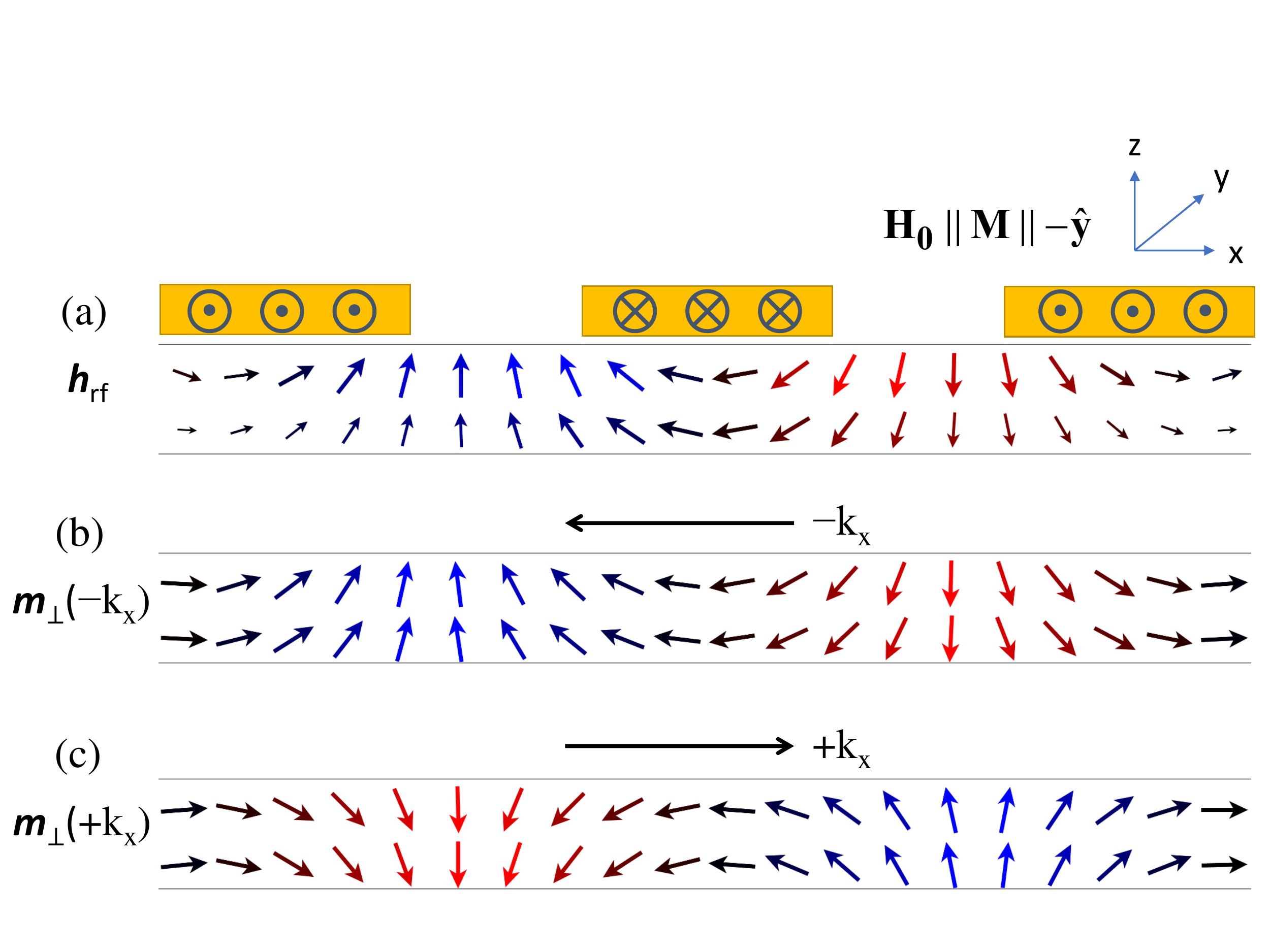}
    \caption{(a) Oersted field from microwave signal applied to a coplanar waveguide (CPW), dynamic magnetization map of a spin waves propagating at (b) the $-x$ direction, and (c) the $+x$ direction. All of the figures show only time-varying component of magnetic field or magnetizations.}
    \label{fig_cpw}
\end{figure}

Aside from tailored spin wave dispersion, a practical approach for implementing nonreciprocal magnon propagation is given by specific geometries of antenna that can selectively excite a desired chirality of spin waves so that the antenna preferentially excites a spin wave propagating in one way compared to opposite direction~\cite{Chen2019,Deorani2014,Shibata2018}. For example, an Oersted field generated from a current through a thin wire placed on top of a film has a definite chirality, i.e., the Oersted field rotates clockwise or counter-clockwise as one moves perpendicular to the wire in the film and this chirality will be reversed if one moves the wire from above the film to below the film. Therefore, the Oersted field excites only one set of spin waves with the same chirality, or the same propagating direction, much more efficiently compared to the other~\cite{Deorani2014}. The same mechanism also applies to the coplanar waveguide (CPW) geometry~\cite{YiLi2023}. Fig.~\ref{fig_cpw}(a) shows a case of CPW where the time-varying Oersted field has the same chirality as that of time-varying component of magnetization of a spin wave propagating at the $-x$ direction as one can see at Fig.~\ref{fig_cpw}(b), i.e., both vectors rotate counter-clockwise as one sweeps the coordinate to the $+x$ direction. A spin wave propagating in a $+x$ direction has opposite chirality as one can see in Fig.~\ref{fig_cpw}(c). Therefore, the CPW excites more efficiently a spin wave propagating to the $-x$ direction than a spin wave propagating in the $+x$ direction.

There are other systems supporting wave nonreciprocity which are not pure spin wave nonreciprocal devices. One example are systems using magnon-phonon interaction~\cite{Kuss2021,Verba2019,Kuss2020,Xu2020,Tateno2020,Kuss2021PRApp,Shah2020,Hernandez2020,Sasaki2017,Maekawa1976,Puebla2022,Matsumoto2022,Kuss2023,Bas2022}. A widely used mechanism is based on Rayleigh surface acoustic waves (RSAWs), which are a kind of circularly polarized wave with opposite chirality in opposite propagating direction. These RSAWs can excite FMR only when the RSAWs have the same chirality as that of the precessing magnetic moments in FM. This chirality selectivity is known as magneto-rotation coupling~\cite{Xu2020,Maekawa1976}. By using the magneto-rotation coupling, a system can have wave nonreciprocity even if its magnetic part does not have nonreciprocity by itself. 

There are other systems using magnon-phonon interaction, where the magnetic part already has nonreciprocity due to interlayer dipolar interaction or interfacial DM interaction as described above~\cite{Kuss2021,Verba2019,Kuss2020,Shah2020,Matsumoto2022}. One notable feature is that some studies reported very high amplitude nonreciprocities, for example, one paper reported $100\%$ isolation~\cite{Xu2020} in which the magnetic part does not have inherent spin wave nonreciprocity, and another paper reported 48.4 dB isolation~\cite{Shah2020} where the magnetic part has a definite spin wave nonreciprocity due to interlayer dipolar interaction from an antiferromagnetically coupled bilayer structure. 

Another feature to note is that there are well-established methods to achieve low insertion loss ($\sim 1$ dB) transduction from microwave to surface acoustic wave (SAW)~\cite{Su2022}. To achieve an insertion loss better than 3 dB, one needs transducer geometries which allow unidirectional propagation of SAW because bidirectional loss already has 3 dB due to its bidirectionality. These unidirectional transducing scheme has been achieved in multiple ways, such as, single-phase unidirectional interdigital transducers (SPUIDT)~\cite{Hartmann1982,Satoh2005}. A low insertion loss transduction is expected to be an essential feature to establish a nonreciprocal quantum state transfer channel which can be used in noise-resilient quantum information devices~\cite{Xiang2017,Vermersch2017}. One study used a unidirectional interdigitated transducer (UDT) to establish a phonon-mediated quantum state transfer and remote qubit entanglement~\cite{Dumur2021} without using magnetic materials. One can expect that a much higher isolation can be achieved and thereby better noise protection if one uses both a combination of UDT and a nonreciprocal magnetic system. Since these systems have no bulky element in contrast to typical microwave system, one can fabricate such integrated quantum information devices on a single chip. This possibility of miniaturization may be able to contribute to realization of high-density, multi-qubit quantum computing processors.

One interesting thing to note is that two different mechanisms of nonreciprocity can interfere ‘constructively’ or ‘destructively’. This depends on how a system is configured. For example, a recent theoretical study expects that a multi-layer system with both interlayer dipolar interactions and interfacial DM interactions is able to have an enhanced nonreciprocity compared to that of interfacial DM interaction alone~\cite{Franco2020}. We speculate that there are many other ways to integrate multiple mechanisms of nonreciprocity so that the mechanisms interfere ‘constructively’, resulting in systems with enhanced nonreciprocity.

\section{\label{sec:magnon-qis}Magnons in hybrid quantum devices}

After optimizing the coherence and non-reciprocity of magnons in ideal candidate materials, the ultimate goal is to coherently couple them to other quantum modules and even qubits to form a hybrid quantum system for faithful quantum information processing. In the past decade many research groups have established coupling between magnons and microwave photons in a cavity~\cite{Tabuchi2015,Zhang2015,Zhang2016SciA,Goryachev2014,Kostylev2016,Match2019,Harder2017} or a coplanar circuit structure~\cite{Bhoi2014,Bhoi2017,Zhang2014,Li2019}. In the work by Tabuchi et al.~\cite{Tabuchi2015}, coherent coupling between a magnon and a superconducting qubit mediated by a microwave cavity has been demostrated. This breakthrough opens possibility for magnons to transfer quantum states in quantum information systems. In this perspective, we will discuss two particular examples: remote magnon-magnon coupling via microwave photon and using non-reciprocal magnons for quantum state transduction.     

\subsection{Remote magnon-magnon coupling}

\begin{figure*}
    \centering
    \includegraphics{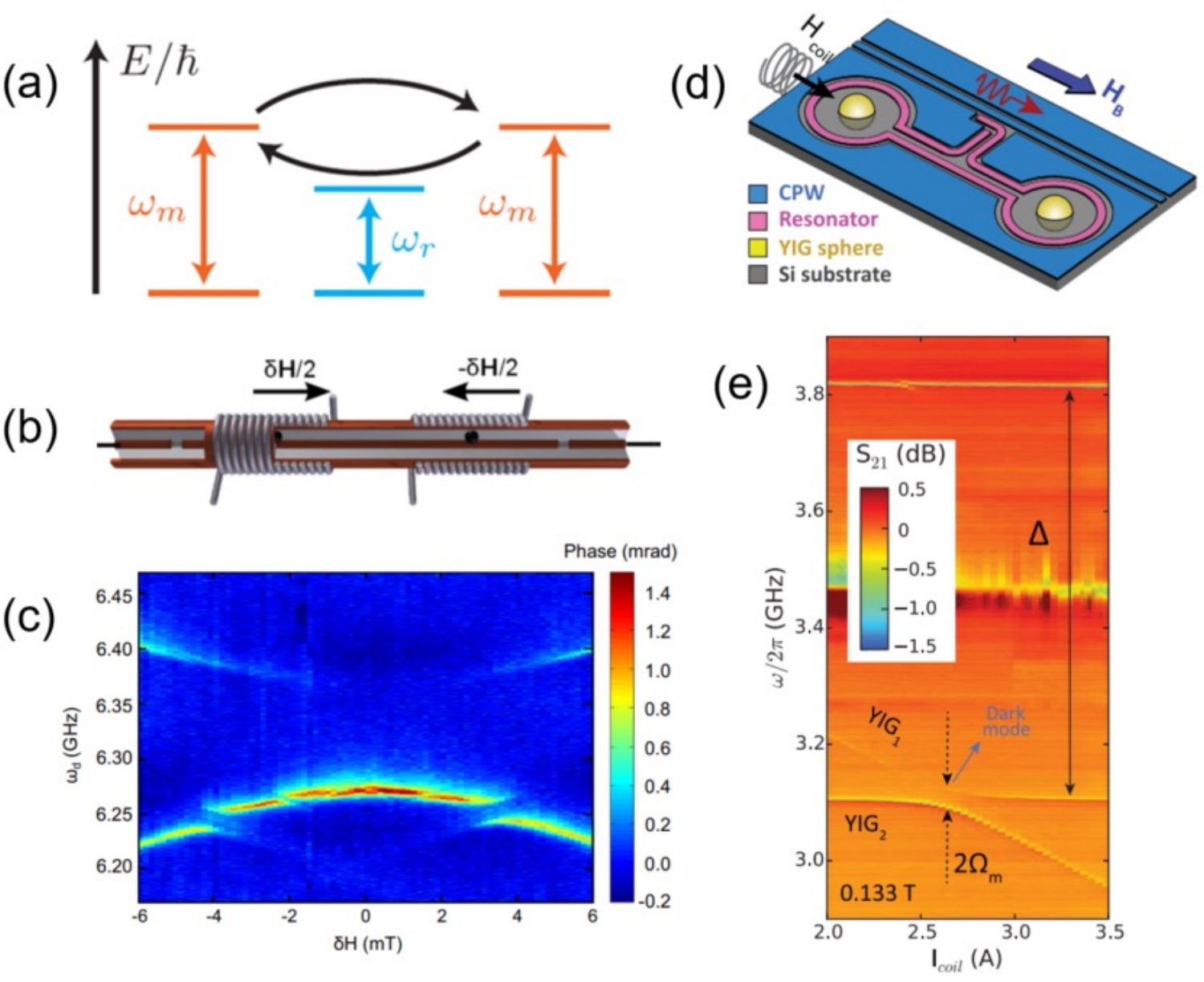}
    \caption{(a)Schematics of magnon-magnon coupling ($\omega_m$) mediated by a photon cavity ($\omega_r$) in the dispersive regime. (b) Cavity design and YIG sphere placement of a coaxial transmission line cavity. (c) Strong coupling of the two YIG spheres mediated by the coaxial transmission line cavity in the dispersive regime. (d) Cavity design and YIG mounting of a NbN superconducting coplanar resonator. (e) Strong coupling between the two chip-embedded YIG spheres with the magnon modes detuned from the superconducting resonator by $\Delta/2\pi=0.7$ GHz. (b) and (c) are adapted from Ref.~\cite{Lambert2016}. (d) and (e) are adapted from Ref.~\cite{Li2022}}\label{fig_qis1:m-m_coupling}
\end{figure*}

Besides coupling magnons to microwave photons and qubits, ongoing efforts of hybrid magnonics also include coupling of multiple remote magnonic resonators aiming to explore remote magnon-magnon entanglements~\cite{Rusconi2019,Ren2022} and implement non-Hermitian magnon-magnon interactions. In particular, remote magnon-magnon interaction can be achieved by dispersive coupling [Fig.~\ref{fig_qis1:m-m_coupling}(a)], where the two magnonic resonators are coupled via exchanging virtual photons through a mutually coupled photon cavity, which is far detuned from the magnon mode, or $|\omega_r - \omega_m| \gg g_{m-r}$, where $\omega_r$ is the resonator frequency, $\omega_m$ is the magnon frequency, and $g_{m-r}$ is the magnon-photon coupling strength. This technique has been commonly used in qubit operation in order to minimize the additional loss from the photon cavity.

The first demonstration of remote magnon-magnon coupling was done by Lambert et al.~\cite{Lambert2016}, where two single-crystal YIG spheres with a diameter of 1 mm were placed within a coaxial transmission line cavity [Fig.~\ref{fig_qis1:m-m_coupling}(b)]. By exciting the second harmonic of the standing wave mode in the cavity, the two YIG spheres are simultaneously coupled to the microwave photon mode with maximal strength. In addition, two coils were wrapped around the transmission line in order to provide detuning magnetic field $\delta H$ and control magnon-magnon interaction. Shown in Fig.~\ref{fig_qis1:m-m_coupling}(c), clear mode splitting is measured between the two YIG spheres when $\delta H=0$, meaning when the two YIG magnon modes are degenerate in frequency. Here the cavity resonance frequency is $\omega_r/2\pi=7\ \text{GHz}$, which is well detuned from the two degenerate magnon mode of $\omega_m/2\pi=6.3\ \text{GHz}$.

Recently, Li et al. have demonstrated similar results with a much smaller system~\cite{Li2022}. By using a superconducting coplanar resonator, the microwave cavity can be made with micron-meter-wide signal lines, which allows for convenient circuit design and device integration on a chip. In addition, a miniaturized microwave cavity enables smaller effective volumes and leads to larger microwave coupling efficiency to magnetization, meaning that the system can be used to couple to smaller magnetic systems along with higher coupling strength. In the work from Li et al, two smaller YIG spheres with a diameter of 0.25 mm were embedded in two etched sockets on the Si substrate supporting the NbN superconducting resonator. The superconducting resonator was designed with two circular antenna to provide a nearly uniform microwave field onto the YIG spheres at the center [Fig.~\ref{fig_qis1:m-m_coupling}(d)]. Furthermore, a local NbTi superconducting coil has been placed next to one sphere for controlling the magnon-magnon frequency difference. Strong magnon-magnon coupling can be clearly observed in Fig.~\ref{fig_qis1:m-m_coupling}(e), with the superconducting cavity frequency ($\omega_r/2\pi=3.8\ \text{GHz}$) being detuned from the degenerate magnon frequency ($\omega_m/2\pi=3.1\ \text{GHz}$).

The remote coupling of magnonic resonators offers a new path to implement hybrid magnonic networks with tunable magnonic interactions such as band structures~\cite{Li2020}. Compared with magnonic crystals that are based on propagating magnons, the band structure of a hybrid magnonic network can be controlled by modifying one or a few magnonic resonator nodes, offering more flexibility of control. The main challenges are the scalability and readout. As more magnonic nodes are involved, there is also the need to incorporate more controlling coils for the current schematic, which will complicate the circuit design. Also it is desired to access the magnon status of each individual magnonic node. The two examples above rely on the readout from the cavity, which will be decoupled from the hybrid magnonic mode with certain phase relationships (dark mode). It is thus desirable to have one designated readout antenna for each magnonic node, similar to the readout of multiple entangled qubits. To replace the use of a local coil, we also envision the use of magnetic thin-film devices, where their frequencies can be modified with spin torques, electric fields or circuit embedded current lines providing an Oersted field.

\subsection{\label{sec:qst}Magnons for non-reciprocal quantum state transfer}
\subsubsection{Role of nonreciprocity in quantum state transfer }

Quantum state transfer (QST) mediated by propagating wavepackets is a promising way to distribute quantum information between distant qubits in so-called cascaded quantum systems, where the output of one qubit forms the input of another~\cite{Carmichael1993, Gardiner1993}. In this scenario, one can use temporal modulation of stimulated emission and absorption to realize deterministic QST with near-unit efficiency~\cite{Cirac1997}. This method can readily be used to realize remote entanglement, and we can thus envision the realization of complex, distributed quantum networks. QST through propagating wavepackets has the following benefits: 1) no static hybridization between qubits; 2) no requirements on length of transmission line channels. This is in contrast to quantum-bus type mediation of interactions which are mediated through standing waves – a common approach for mediating interactions in closed systems~\cite{Sorensen1999, Majer2007}.

Realization of QST in this way relies on directional elements. First, they aid in avoiding standing waves in finite-length transmission lines, thus ensuring the ‘propagating wavepacket’ scenario. Second, they ‘route’ wavepackets between emitters (senders) and absorbers (receivers), which makes it also highly appealing to have in-situ tunable directionality.

Chiral networks have been predicted to have interesting features that may be useful for studying open quantum systems, and for suppressing noise and loss. For one, driving cascaded systems has been predicted to enable the realization of dissipatively stabilized remote-entangled states~\cite{Stannigel2012}. In addition, chiral coupling of unidirectionally traveling wavepackets to qubits could allow remarkably noise-resilient distribution of quantum states and entanglement~\cite{Xiang2017, Vermersch2017}.

\subsubsection{State of the art and current approaches}
QST through traveling signals is particularly interesting in superconducting microwave quantum circuits. These systems can combine low intrinsic photon loss with excellent temporal control over qubit-photon coupling (for a review, see Blais et al.~\cite{Blais2021}). At the same time, the need for distributing quantum information across multiple devices has been recognized as an important need for building large-scale quantum devices from the bottom up~\cite{Niu2023,Awschalom2021PRX,Jiang2007,Kimble2008,Monroe2014,Chou2018,LaRacuente2022,McKinney2022}, making QST a very timely topic. 

Several experiments have, in recent years, demonstrated deterministic QST and entanglement with decent fidelities using the above approach~\cite{Axline2018,Campagne2018,Kurpiers2018,Magnard2020}. In these experiments the directionality was provided by commercial microwave circulators. The use of these elements and associated connectors has, however, resulted in levels of photon loss that make scaling beyond two nodes difficult to impossible. In addition, these circulators have fixed directionality, preventing the construction of networks with high connectivity.

One way to overcome these challenges is to realize non-reciprocal elements using Josephson circuits. The introduction of magnetic flux through loops containing junctions allows breaking Lorentz symmetry and the realization of devices in which microwave excitations are emitted or scattered nonreciprocally~\cite{Koch2010,Kamal2011}. Importantly, such Josephson circuits can be expected to be lossless, and magnetic flux tuning, which is a well-established tool in quantum circuits, allows in-situ control over the directionality. Recent theoretical~\cite{Guimond2020,YXZhang2021,Kerckhoff2015,Gheeraert2020,Chapman2019} and experimental~\cite{Abdo2021,Kannan2023,Chapman2017,Rosenthal2017} works have explored Josephson-based directional devices and how to construct multinode quantum networks from them. These results show great promise for realizing high-fidelity state transfer.

\subsubsection{Potential for magnon-mediated QST}

\begin{figure}
    \centering
    \includegraphics[width=8.6cm]{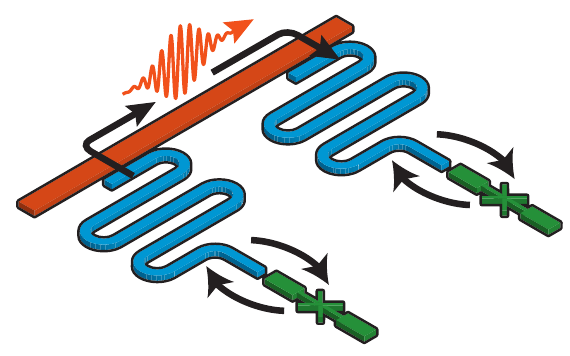}
    \caption{Schematic of unidirectional quantum state transfer with nonreciprocal magnetic waveguide. The system consists of two transmon qubits (green), a nonreciprocal magnonic waveguide (orange), and two intermediary superconducting resonator (blue).}\label{fig_qis2:QST}
\end{figure}

While the above approach for realizing synthetic nonreciprocity is highly appealing, there are downsides. In particular, the fact that many of the proposed devices are fairly narrow-band and that they require multiple control channels (rf or dc connections) to tune and/or operate raises the question whether simpler passive devices are possible to be integrated into quantum circuits in a similar way. The recent advances in hybrid-magnon devices could potentially show the way toward such elements as we described in Sec.~\ref{sec:nonreciprocal}. The operating band of hybrid magnonic devices is easily adjustable with varying magnetic field strength and these hybrid magnonic systems do not need multiple control channels but rather just a rf channel. Fig.~\ref{fig_qis2:QST} shows a schematics of a unidirectional quantum state transferring system which could realize an above-mentioned noise-resilient quantum information device.

In parallel, there are studies of magnon-assisted nonreciprocal microwave photon propagation in the context of chiral cavity quantum electrodynamics~\cite{YYWang2021,Yu2020,XueFengZhang2020,Zhu2020}. For example, for a ferrimagnetic-cylinder-loaded Y-shaped cavity, it was demonstrated that this device acts as a microwave circulator~\cite{YYWang2021}. The main mechanism was due to the fact that two almost-degenerate chiral magnon-polariton modes with opposite propagation direction were canceled at one port and not canceled at the other port. A study of a more complicated system consisting of a $5 \times 5$ array of cavities, where four of them include ferrimagnetic sphere inside, has been reported~\cite{Owens2022}. It turned out that the system has bulk modes and chiral unidirectional surface modes with topological properties. Also, a ferrimagnetic sphere-loaded X-shaped cavity showed nonreciprocal microwave propagation~\cite{YPWang2019}. This system has both coherent and dissipative coupling between magnon and cavity photons and the dissipative coupling parameter has a phase difference between two oppositely propagating microwave photons so the system has nonreciprocal propagation character.

Devices based on currently known hybrid-magnon circuits might need external magnetic fields, which is not readily compatible with many quantum device platforms, in particular superconducting qubits. However, there have been significant advances in field-compatible qubits, such as spin-qubits or variations of superconducting devices~\cite{Krause2022}, that suggest this may not be an insurmountable roadblock.

\section{\label{sec:conclusion}Conclusions}

Hybrid quantum systems are gaining increasing interest for their potential in coherent information processing because one can expect to exploit combined advantages from each hybridizing component. In this context, magnon-based hybrid systems, known as hybrid magnonics, possess unique advantages from magnons, quanta of spin waves in a magnetic material. Magnons can hybridize to a variety of other excitations, especially microwave photons, in the strong coupling regime. The intrinsic non-reciprocity opens the potential of magnons for unidirectional quantum state transfer. The characteristic dispersion relation of magnons allows for miniaturized microwave magnonic devices. All these suggest that magnon-based hybrid quantum systems are very promising candidate platforms for quantum information science.                

To achieve robust hybrid-magnon quantum devices, it is essential to have optimized material candidates with low magnetic damping such that the coherence of the hybrid systems can stay as long as possible. Both the ferrimagnetic insulator YIG and metallic ferromagnets show promising potential towards this end. Considering the on-chip integrability and the relevance for quantum information applications, we are especially interested in the damping behavior of these materials in a thin film structure and at low environmental temperatures. For YIG, due its exceptionally low intrinsic damping, prior effects should be focused on reducing external damping sources that can come from surface imperfections, chemical impurities, and substrate (GGG) effects with more advanced experimental techniques. For metallic ferromagnets, magnon-electron scattering dominates the damping in most situations and complicates its temperature dependence. First-principles calculations of the electronic structure can provide guidance for searching possible promising materials. Engineering electron energy bands by tuning alloy compositions, chemical doping and structure amorphization provide several pathways for optimization. Recently increased interest in 2D van der Waals magnets and topological magnetic semi-metals may provide other new opportunities towards this goal.          

Another advantage of magnon-based hybrid quantum systems is that ferro/ferrimagnetic systems naturally have time reversal symmetry breaking due to the presence of non-zero net magnetization. This feature allows one to generate and engineer wave nonreciprocity in ferro/ferrimagnetic systems. We have reviewed various mechanisms which result in spin wave nonreciprocity. A nonreciprocal hybrid magnetic system has unique properties including tunable operating frequency by changing applied field and it is miniaturizable compared to currently used microwave devices (e.g., isolators or circulators). As described in Sec.~\ref{sec:nonreciprocal}, these nonreciprocal magnetic systems may be able to provide noise isolation on the chip level. Also, as described in Sec.~\ref{sec:qst}, there are protocols of quantum state transfer in the presence of background noise~\cite{Xiang2017, Vermersch2017}, which need a nonreciprocal channel between qubits. We think nonreciprocal magnetic systems may be able to provide suitable functionality to establish such noise-resilient quantum information devices with the field compatible qubits~\cite{Krause2022}. 

With optimized low damping and non-reciprocity in ideal material candidates, coupling them to hybrid quantum systems for coherent information processing is the ultimate goal. The last decade has witnessed wide-spread effort and significant progress towards this goal. Two particular relevant aspects have been discussed in this perspective: remote magnon-magnon coupling mediated by microwave photons in a superconducting coplanar resonator and the possibility for magnons assisting with non-reciprocal quantum state transfer. Furthermore, there are many additional fundamental opportunities worth to be explored, such as to manifest hybrid magnonic quantum information devices on a single chip, including low magnetic damping, low-insertion-loss transduction between microwave photon and magnon, large enough magnon amplitude nonreciprocity, and magnetic field resilient qubit development. The field is attracting continuous interest from the scientific community. With on-going, and more and more future efforts, we believe that magnon-based hybrid systems will be one of the promising platforms for faithful coherent quantum information processing.

\begin{acknowledgments}
The work was supported by the U.S. DOE, Office of Science, Basic Energy Sciences, Materials Sciences and Engineering Division, with all parts of the manuscript preparation supported under contract No. DE-SC0022060.
\end{acknowledgments}

\bibliography{introduction, dp, nr, qis_pt1, qis_pt2}

\end{document}